# Novel Josephson Junction Geometries in NbCu bilayers fabricated by Focused Ion Beam Microscope

R. H. HADFIELD, G. BURNELL, P. K. GRIMES, D.-J. KANG, M. G. BLAMIRE

IRC in Superconductivity and Department of Materials Science, University of Cambridge, U.K.

Pembroke Street Cambridge, CB2 3QZ, United Kingdom

Corresponding Author:
Robert Hadfield, Department of Materials Science, University of Cambridge, U.K., Pembroke Street, Cambridge, CB2 3QZ, United Kingdom. Tel: +44 1223 334375. Fax: +44 1223 334373, E-mail: rhh20@hermes.cam.ac.uk

**Abstract—We explore novel junction configurations as an extension of our established Focused Ion Beam–based low $T_C$ SNS Junction fabrication technique[1]. By milling a circular trench (diameter 1 μm, width 50 nm) in a 125 nm Nb 75 nm Cu bilayer we define a superconducting island connected to the bulk of the film by a normal metal barrier and entirely enclosed in-plane by the superconducting film. The properties of this Corbino geometry Josephson junction can be probed by depositing an insulating layer over the device and drilling a 0.3 μm diameter hole down to the island to allow a Nb via to be deposited. Behavior of such devices has been studied in a Helium bath at 4.2 K. An SNS-like current voltage characteristic and Shapiro steps are observed. It is in terms of magnetic field behavior that the device exhibits novel characteristics: as the device is entirely enclosed in type II superconductor, when a magnetic field is applied perpendicular to the plane of the film, only quantized flux can enter the junction. Hence as applied magnetic field is increased the junction critical current is unchanged, then abruptly suppressed as soon as a flux quantum enters.**

**PACS Codes and Keywords: 74.50+r Josephson effects; 74.80.Fp Superconducting junctions; 85.25.Cp Josephson devices.**

# 1. Introduction

In a conventional Josephson junction magnetic flux can enter progressively through the edges of the junction. As a result the phase difference between the electrodes varies across the width of the junction, leading to a suppression of the junction critical current. In the simplest 2-dimensional geometry the critical current versus magnetic field, $I_C(B)$, follows a modulus of a sinc function dependence, analogous to the diffraction pattern of monochromatic light due to a single slit in the Fraunhofer regime [1]. The first minimum occurs when one flux quantum $\Phi_0$ is present in the barrier. Consider now the case of a Josephson junction in a Corbino geometry (this geometry is used for example in studies of the Quantum Hall effect, as it allows edge effects to be neglected [2]). The barrier region is now entirely enclosed in a loop of superconductor: magnetic flux is only permitted to enter in single quanta. The critical current versus magnetic field dependence should now abruptly switch from the maximum value to zero as flux enters, resulting in a top hat $I_C(B)$ pattern (Fig. 1).

Using our established SNS planar bridge fabrication technology we have succeeded in realizing such a device. Our conventional fabrication technique involves milling a 50 nm wide trench in the upper superconducting layer of a NbCu bilayer track using a Focused Ion Beam microscope in order to achieve weak coupling [3]. Using this technique devices with resistively-shunted junction model-type current-voltage characteristics, critical current, $I_C$ 1 mA per micron track width and characteristic voltage, $I_C R_N$ ~50 µV at 4.2 K are routinely obtained [4]. In order to realize the new device architecture we mill a circular trench in the bilayer. An insulating layer of silica is then deposited over the structure. A hole is then drilled through the silica using the FIB. A Nb via is then deposited, facilitating electrical contact to the central island.

# 2. Fabrication Technique

An oxidized silicon substrate is patterned for lift-off by optical photolithography. A NbCu bilayer (125 nm Nb on 75 nm Cu) is deposited in an ultra-high vacuum magnetron sputtering system. Hence a standard pattern of microscopic tracks is defined. The sample is then transferred to the FIB system (FEI Inc. FIB200). An initial calibration trench is milled through on a redundant track whilst making a 4 point *in situ* resistance measurement [5]. From this measurement the milling rate per unit length of trench can be deduced. A circular trench (radius 0.75 µm, trench width 50 nm) is then milled in the bilayer 80% of the way through the upper Nb layer. The remaining Nb in the bottom of the trench carries a Josephson current in parallel with the Cu beneath, leading to an enhanced $I_C$ at 4.2 K [4]. In addition isolation cuts are made in the basic track structure (Fig. 2). The sample is then lift-off patterned a second time and a layer of silica (300 nm thick) is deposited. The sample is then returned to the FIB. A hole of 0.3 µm radius is drilled through the silica layer down to each circular island. From a measurement of the stage current it is clear when the insulating layer is breached. After a further lift-off patterning step a 300 nm thick Nb layer is deposited by UHV magnetron sputtering. A section through a finished device prepared and viewed in the FIB at 45º tilt is shown in Fig. 3.

## 3.  Measurements and Discussion

Quasi-static current-biased measurements were carried out using a dip-probe at 4.2 K. All the devices measured in a Helium bath at 4.2 K had small critical current ($I_C$ ~10 µA) with a resistance in series (~1 Ω; attributable to the contact resistance between the island and the top electrode).  A current-voltage characteristic with and without microwaves (13.0 GHz) at 4.2 K is shown in Fig. 4.  The appearance of Shapiro steps under applied microwave power comfirms this critical current to be a Josephson current.  The observed critical current is two orders of magnitude lower than expected from a conventional SNS junction of this trench aspect ratio and overall width. (Circular junction circumference 4.7 µm: $I_C$ =15 µA at 4.2 K; conventional SNS planar junction trench length 1.5 µm: $I_C$ = 1000 µA at 4.2 K.

The $I_C$(B)  behavior of the devices in question was probed by applying a magnetic field perpendicular to the plane of the film.  The devices were cooled whilst shorted in zero field.  The external field was increased to 50 mT, decreased to –50 mT and then returned to zero.  The resulting $I_C$(B) pattern is shown in Fig. 5.  The device exhibits an abrupt switching in the critical current, resulting in a top hat shape.  The switching occurs at approximately the same field in each case as the magnitude of the field is increased from zero.  The trapped flux quantum is only annihilated when the sign of the applied field is reversed, leading to the apparent hysteresis.  At present exotic fluxon modes or Fiske resonances within the barrier have not been observed.  This is a consequence of the low Q-factor of the normal metal barrier.

The effect on the junction $I_C$ of a flux vortex at various positions was modelled by considering the effect of the vortex field on the critical current in a circumferential element of the junction and integrating.  The results of the numerical calculation are shown in Fig. 6.  As the vortex approaches the region of the junction the $I_C$ falls to zero.  It is of interest to note that if the penetration depth approaches the radius of the junction, the observed critical current is less than the theoretical maximum.  This indicates that the trapping of flux quanta in the centre of the island (which is probably rich in pinning sites as a consequence of the milling of the via hole) will lead to an appreciable lowering of the observed critical current.  A more plausible explanation for the drastic reduction in $I_C$ however may simply be that the significant series resistance due to the via contact leads to heating and the effective temperature of the device is greater than 4.2 K.  An improved via technology would lead to a lower contact resistance, hence less heating and an enhanced $I_C$.

# 4.    Conclusion

In this study we have succeeded in realizing a Josephson junction in a Corbino geometry. The junction region is enclosed in a superconducting loop, so magnetic flux can only enter the junction region as single quanta, leading to an abrupt switching in the critical current from its maximum value to zero.  The observed critical current is significantly lower than that expected for a linear junction of equal barrier length.  This may either be a consequence of flux trapping in the centre of the island or due to heating arising from the via contact resistance.  It would also be possible to create such a device in a film of cuprate superconductor.  A suitable technique for creating such a junction would be with Focused Electron Beam Irradiation (FEBI) [6].

## ACKNOWLEDGMENTS

This work was sponsored by the U.K. Engineering and Physical Sciences Research Council (EPSRC) the U.K. Institute of Physics, the Armourers and Brasiers Company and British Alcan Aluminium.  R.H.H. would like to thank Dr Edward Tarte  for useful discussions, and Moon-Ho Jo and Dr Jose Luis Prieto for technical assistance.

**FIGURE CAPTIONS**

Figure 1: Schematic contrasting the magnetic field response of a conventional planar SNS junction with that of a Corbino geometry Josephson junction.

Figure 2: FIB image of a sample after initial nanopatterning. Circular trenches have been milled for devices (total mill times 15, 18, 21 and 24 s left to right).

Figure 3: FIB image of device architecture (sample sectioned and viewed at 45° tilt in FIB).

Figure 4: Current-Voltage (I-V) characteristic of a device at 4.2 K (series resistance subtracted). Microwaves applied at 13.0 GHz

Figure 5: Critical current versus magnetic field ($I_C(B)$) for a device at 4.2 K. Magnetic field is applied perpendicular to the film. The numbers and arrows indicate the sequence and direction of the magnetic field sweep. $I_C$ is extracted using a voltage criterion resulting the apparent non-zero minimum value.

Figure 6: Numerical predictions of the variation of junction of critical current with vortex position for various values of penetration depth/junction radius. The model breaks down when the vortex nears the barrier region. When the vortex is in the barrier (normalized distance =1) we of course expect $I_C$ to fall to zero.

## FIGURES:

Conventional SNS junction: magnetic flux
enters through sides of junction

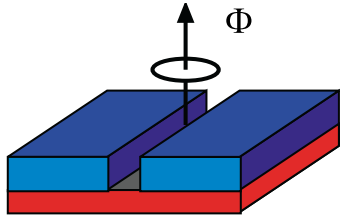

Magnetic field response of critical current:

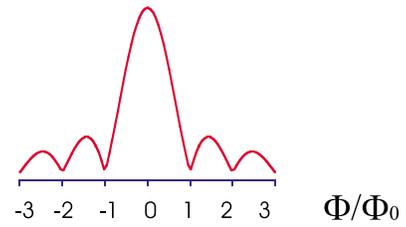

Corbino geometry junction: magnetic flux
can only enter as single quanta

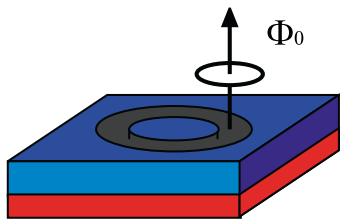

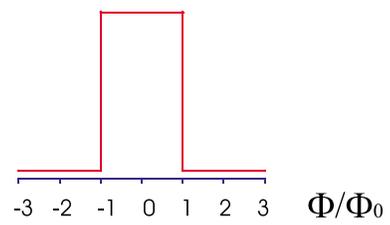

Figure 1

Circular Trenches
(mill times 15, 18, 21, 24 s left to right)

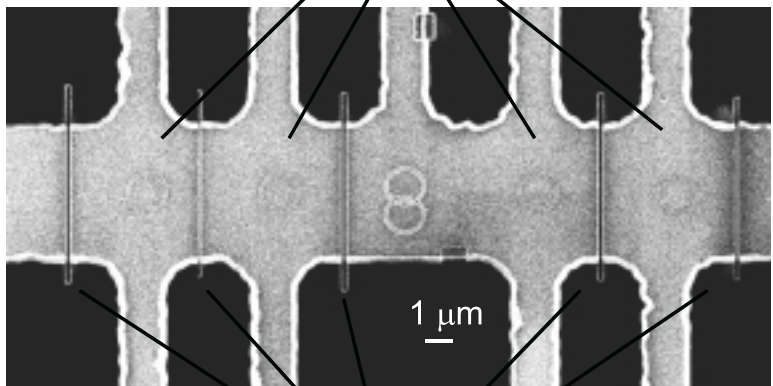

1 μm

Figure 2

Isolation Cuts

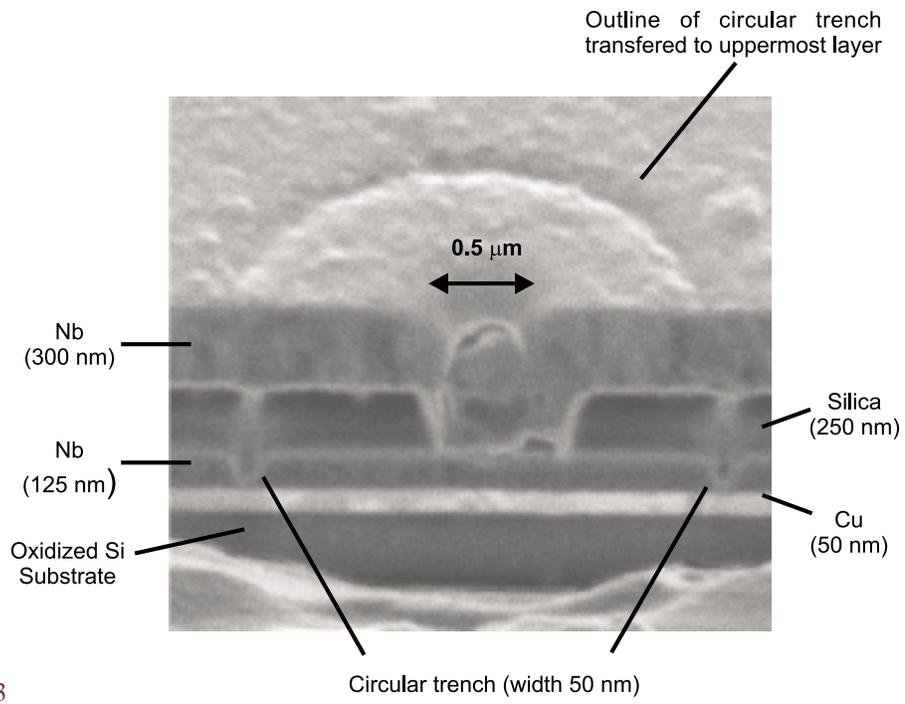

Outline of circular trench
transfered to uppermost layer

**0.5 μm**

Nb
(300 nm)

Nb
(125 nm)

Oxidized Si
Substrate

Silica
(250 nm)

Cu
(50 nm)

Figure 3

Circular trench (width 50 nm)

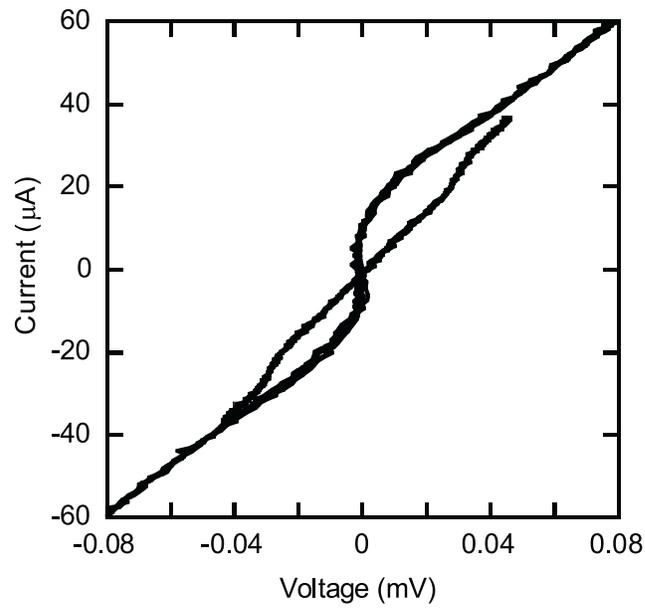

Figure 4

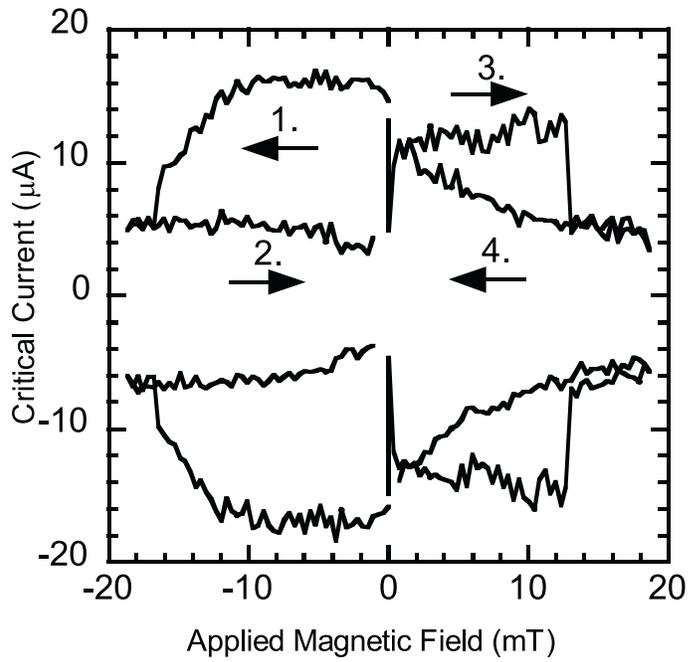

Figure 5

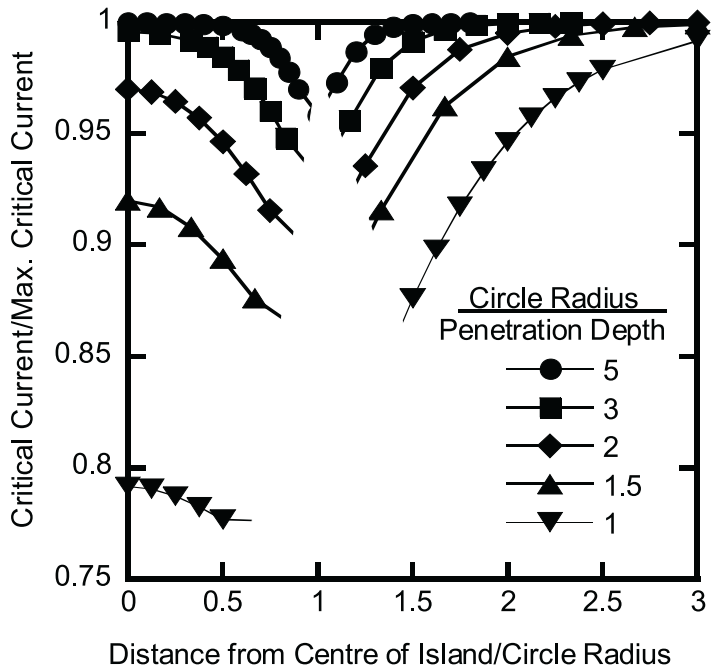

Figure 6